
\magnification=\magstep1
\baselineskip=18pt
\overfullrule=0pt
\nopagenumbers
\footline={\ifnum\pageno>1\hfil\folio\hfil\else\hfil\fi}
\font\twelvebf=cmbx12

\rightline{CU-TP-547}
\rightline{January 1992}
\vskip .8in
\centerline{\twelvebf The Secret Chern-Simons Action for the Hot Gluon
Plasma}
\vskip .8in
\centerline{\it Ravit Efraty and V.P.Nair}
\smallskip
\centerline{Physics Department, Columbia University}
\centerline{New York, NY 10027}
\vskip 1in
\centerline{\bf Abstract}
\smallskip
\noindent We show that the generating functional for hard thermal loops
with external gluons
in QCD is essentially given by the eikonal for a Chern-Simons gauge
theory.
This action, determined essentially by gauge invariance arguments,
also gives an efficient way of obtaining the hard thermal loop
contributions without the more involved calculation of Feynman
diagrams.
\vskip 1.5in
\noindent This research was supported in part by the U.S. Department
of Energy.
\vfill\eject

The Chern-Simons (C-S) action made its appearance in physics literature
over
ten years ago as a mass term for gauge fields in three dimensions $^1$.
Studies
since
then have revealed many interesting properties of this action. The
Abelian
version can be used for spin transmutation, converting spin zero bosons
into
anyons, for example $^2$. The correlators of Wilson lines in a pure
C-S theory are related to the the polynomial invariants of knot theory $^3$.
Pure
C-S theory is also closely related to conformal field theory and the
Wess-Zumino-Witten (WZW) action in two dimensions $^{3,4}$. Also actions
related
to the
C-S action can be used for self-dual  gauge fields and integrable
systems $^5$.
Finally there is an intriguing class of vortex solutions in
spontaneously
broken
C-S theory $^6$. However, despite  this bounty of
interesting results there have not been many realistic physical systems
for
which
the C-S action is relevant. In this letter we show that the C-S action,
more
precisely its eikonal, is part of the effective action for describing
the gluon plasma in Quantum Chromodynamics (QCD). This action,
determined essentially by gauge invariance requirements, gives an
efficient way of obtaining the hard thermal loop contributions, without
having to calculate the corresponding Feynman diagrams. The C-S
connection is
particularly interesting in view of the possibility of producing the
quark-gluon
plasma in heavy ion collisions in the near future.

We consider QCD at temperatures well into the deconfinement phase; i.e.
we have a `hot' plasma of gluons. The effective action mentioned above
is more precisely
defined as follows. Pisarski has shown that a partial resummation of
Feynman
diagrams in thermal QCD is necessary to obtain gauge invariant results,
for
example for the gluon damping rate in the plasma $^7$. The resummation
amounts
to the following. We calculate the one-loop diagrams of
thermal QCD; the relevant kinematical regime corresponds to the
loop momentum being much larger than the external momenta.
These are
the so called hard thermal loop contributions. For these, the external
momenta
are typically of the order of $gT$ where $g$ is the coupling constant
and $T$
is
the temperature; the loop momentum being hard, i.e. at least
of the order of $T$,
is the
region of interest. The leading contributions are proportional to
$T^2$.
The generating functional for hard thermal loops is the
effective action. Thus, once the high temperature contributions of the
hard
thermal loops have been obtained, calculations can be done for any
process
starting from the effective action.  The result is gauge invariant and
consistently accounts for all terms of a given order in coupling
constant $^7$.

Many authors $^{7,8}$ have written down
versions  of this generating functional. The results involve a null
vector
$Q_\mu = (1, \vec Q)$ and integration $\int d\Omega$ over the directions
of the
unit vector $\vec Q$. Diagrammatically, $\vec Q$ arises as follows.
Thermal
loops describe the absorption and emission of particles from the
surrounding
medium or thermal bath. These particles are on mass-shell and thus the
loop
integration is only integration over the momentum three-vector with a
distribution of the Bose-Einstein or Fermi-Dirac form.
Integration over the magnitude
of this
vector is then carried out leaving the angles, described by the unit
vector
$\vec Q$, for the final integration. We shall also need the following
coordinates
$(u,v,\vec x_T), ~u = {1\over 2}Q^\prime \cdot x, ~v = {1\over 2}Q\cdot x,
{}~\vec Q \cdot \vec x_T = 0$, where $Q^{\prime}_\mu = (1, -\vec Q).$ The
components of the gauge field along $Q$ will be denoted by $A^a$, viz.
$A_\mu =
-it^a A_\mu^a$, $Q\cdot A^a = A^a$, $Q \cdot A = A.$  $~\{t^a\}$ are a basis
of the Lie algebra of the group $G$, chosen as hermitian matrices in the
fundamental representation with ${\rm Tr}(t^a t^b)={1\over
2}\delta^{ab}$. Following reference
8, the
generating functional for hard thermal loops with external gluons
has the structure
$$
\Gamma = {CT^2\over 12\pi}
(2\pi \int d^4x~ A_0^aA_0^a  +
\int d\Omega~
W) \eqno(1)
$$
$C=C_q$ for quark loop contributions and $C= C_G$ for gluon-ghost
loop contributions; $C_q$ and $C_G$ are the quadratic Casimirs for the
quark and adjoint representations respectively. We shall not display the
coupling constant $g$ in what follows, as it can be recovered by
$A_\mu \rightarrow gA_\mu$ and an overall ${1\over g^2}$.
The first term in brackets is the well known mass term for the
time-component of the gauge field. The second term may be considered as
what
is
necessary to render $\Gamma$ gauge invariant. The information given by
the
diagrammatic  analysis of hard thermal loops is that it can be written
as $\int
d\Omega ~W$, where $W$ is a functional of $A=Q_\mu A_\mu.$

The condition
for the
gauge invariance of $\Gamma$ in (1) is
$$
\int d\Omega ~\delta W = 4\pi
\int
d^4x ~\dot A_0^a \omega^a \eqno(2)
$$
where $\delta A_\mu = \partial_\mu
\omega
+[A_\mu, \omega].$ Equation (2) is realized by
$$
\delta W = \int
d^4x~ \dot
A^a \omega ^a \eqno (3)
$$
One can check that (3) is indeed the way
gauge
invariance is realized, again by analysis of diagrams. Thus it is
equation (3)
that we must solve. We rewrite this, using the transformation law for
$A^a$,
as
$$ {\partial f \over \partial u} + [A,f] = -{1 \over 2} {\partial A
 \over \partial v} \eqno (4)$$
where $f= {{\delta W}\over {\delta A}}+ {\textstyle {1\over 2}}
A$. Our analysis so far parallels reference 8. However, we shall now solve
(3,4) in terms of the eikonal for the Chern-Simons theory.

We begin by briefly recalling some aspects of the pure Chern-Simons
theory $^{3,4}$.
Consider the C-S action
$$ S = {k \over 4\pi} \int\limits_{R^3} d^3x~{\rm Tr}(a_\mu \partial_\nu
a_\alpha  + {\textstyle{2 \over
3}}a_\mu a_\nu a_\alpha )\epsilon^{\mu\nu\alpha}
\eqno(5)$$
Here $a_\mu$ is the Lie algebra valued gauge potential,
$a_\mu = -it^a a^a_\mu $.
We shall use complex
coordinates $z,\bar z ,~z=x +iy, $
to describe the spatial dimensions. The time-component $a_0$ can be set
to zero
as a gauge choice. The equations of motion then tell us that $a_z ,
a_{\bar z}$
are independent of time but satisfy the constraint
$$
\partial_za_{\bar z} - \partial_{\bar z}a_z + [a_z ,a_{\bar z}] =0.
\eqno(6)
$$
This can be solved for $a_{\bar z}$ as a function of $a_z$, at least as
a power
series in $a_z$. The result is
$$a_{\bar z} = \sum (-1)^{n-1} \int {d^2z_1 \over \pi} \dots {d^2 z_n \over
\pi}
{a_z(z_1, \bar z_1) \dots a_z (z_n, \bar z_n) \over (\bar z - \bar z_1)(\bar
z_1 -
\bar z_2)
\dots (\bar z_n - \bar z)} \eqno(7)
$$
This can be checked easily using $\partial_z {1\over {\bar z}-{\bar
z}'}= \pi \delta^{(2)}(z-z')$.

Define the functional $I[a_z]$ such that
$$
\delta I= {ik\over \pi}\int d^2x {\rm Tr}(a_{\bar z}[a_z]~\delta a_z)
\eqno(8)
$$
Since $a_{\bar z}$ is conjugate to $a_z$,
$ I $ so defined is the action evaluated for the classical motion; thus
it is Hamilton's principal function or the eikonal for the C-S action.
(Recall that the eikonal for one-dimensional particle mechanics is the
integral of $pdx$, where $p$ is the canonical momentum, specified as a
function of $x$ by fixing the energy. We have an analogous situation
where $a_{\bar z}$ is specified in terms of $a_z$ by equation (6).)
We can write $I$ as
$$I = ik\sum {(-1)^n \over n} \int {d^2z_1\over \pi} ...{d^2z_n\over
\pi}~{{\rm Tr} \bigl( a_z (z_1, \bar z_1) \dots
a_z (z_n, \bar z_n)
\bigr) \over {\bar z}_{12}{\bar z}_{23} \dots {\bar z}_{n1}} \eqno(9)
$$
where ${\bar z}_{ij} = {\bar z}_i -{\bar z}_j.$
(This result is fairly standard; for an
elegant way of obtaining and rewriting this, see reference 9.)
$I$ is in fact the WZW action. If we
parametrize
the two-dimensional gauge field as $a_z = -\partial_z U~U^{-1}$
where $U$ is valued in the complexification of the group $G$, (9) can be
written in the more conventional form of the WZW action $^{10}$ as
$I=-ikS_{WZW}$, where
$$
S_{WZW} = {1 \over 2\pi }\int d^2x~{\rm Tr}
(\partial_z U\partial_{\bar z} U^{-1}) -
{i \over 12\pi}
\int\limits_{M^3}d^3x~ {\rm Tr}(U^{-1}\partial_\mu U~U^{-1}\partial_\nu
U~U^{-1}\partial_\alpha U)\epsilon^{\mu\nu\alpha} \eqno(10)
$$
(As usual $M^3 = R^2 \times [0,1]$ with $U(z, {\bar z}, 0)=1,
{}~U(z, \bar z, 1) = U(z,\bar
z).$)
The WZW action is thus the eikonal for the C-S action.
For our discussion below, $a_\mu$ will not be just a two-dimensional gauge
field.
Equation (9) will thus be the more useful form. Also, since $k$ is not
relevant for our discussion, we shall henceforth set it to one.

Returning to equation(4), notice that it is of the form of a zero
curvature condition. We first do a Wick rotation to Euclidian space so that
$2u\rightarrow z,~2v\rightarrow {\bar z},~$
$\partial _u \rightarrow 2 \partial_z,$
$\partial _v \rightarrow 2 \partial_{\bar z}.$
Defining $a_{\bar z} = -f$ and $a_z = {1\over 2} A $, equation (4) is
seen to be identical to (6). Hence the solution for $W$ is given by
$$
W = -{1\over 4 } \int d^4x~A^a A^a -4\pi i I[A/2] \eqno(11) $$
where, from (9) and the definition of $a_z$,
$$I[A/2] = i~\sum {(-1)^n \over n } \int d^2 x_T ~{d^2z_1\over
\pi} \dots {d^2z_n\over \pi}
{}~{1\over 2^n}~{\rm Tr} {\bigl( A(x_1) \dots A(x_n) \bigr) \over
(\bar z_{12} \bar z_{23} \dots \bar z_{n1})} \eqno(12)
$$
The potentials now depend on all four coordinates; however since (4)
does not involve differentiations with
respect to the transverse coordinates
$\vec x_T$, all potentials in (12) have the same argument for these
coordinates.
In other words, the transverse coordinates in (12)
only play the role of parameters on
which the $A$'s depend. Using (11) in (1) we have the generating
functional in terms of the eikonal $I$,
$$
\Gamma = {CT^2\over 12\pi} \left[\bigl(
\int d^4 x~
 (2\pi A_0^a A_0^a
- {\textstyle{1 \over 4}} \int d\Omega~A^a A^a \bigr)
- 4\pi i \int d\Omega~ I[A/2]\right] \eqno(13)
$$

We have not actually calculated Feynman diagrams to arrive at (13).
The only input from a diagrammatic analysis has been the structure
$\int d\Omega~ W.$ However it is easy to check that the $n$-point
functions
calculated from (13) agree with the explicit diagrammatic evaluation
of hard thermal loops.
The $n$-point functions in momentum space are given by
$$ (2\pi)^4 \delta^{(4)} (\sum k_i) ~\Gamma_{\mu_1 \dots \mu_n}^{a_1 \dots a_n}
(k_1 \dots k_n) = \int d^4x_1 \dots d^4x_n~
e^{-i\sum k_i \cdot x_i} ~V^{a_1 ...a_n}_{\mu_1 ...\mu_n}(x_1,...x_n)
$$
$$
V^{a_1...a_n}_{\mu_1...\mu_n}(x_1,...x_n)
 =\left[{\delta^n \Gamma \over \delta
A_{\mu_1}^{a_1}(x_1)
\dots \delta A_{\mu_n}^{a_n}(x_n)}\right]_{A=0} \eqno(14)
$$
The two-point function is given by
$$
\Gamma ^{ab}_{\mu \nu}(x_1, x_2) = \delta^{ab}{C T^2\over 12\pi}
\big[ 4 \pi \delta_{\mu 0} \delta_{\nu 0}
\delta ^{(4)}(x_1-x_2)
$$
$$
-{\textstyle {1\over 2}}\int d \Omega ~
Q_\mu Q_\nu \bigl( \delta ^{(4)}(x_1-x_2) - {\delta ^{(2)}(x_{T_1} -
x_{T_2}) \over \pi (\bar z_1 -\bar z_2)^2} \bigr) \big] \eqno(15)
$$
We need the Fourier transform to obtain the expression in
momentum space. This is straightforward for the first two terms.
For the last term we have an expression of the form
$$
H = \int {d^2z_2 \over \pi}{h(z_2, \bar z_2) \over (\bar z_1 - \bar z_2)^2}
= - \partial_{\bar z_1} \int {d^2z_2\over \pi}
{h(z_2, \bar z_2) \over (\bar z_1 - \bar z_2)}
\eqno (16)
$$
where $h$ is an exponential of the form $\exp i (k_z \bar z + k_{\bar z}
z).$ Using
$\partial_z {1 \over (\bar z - \bar z^\prime)} = \pi \delta ^{(2)}
(z - z^\prime),$
we get
$$
\partial_{z_1} H = -\partial_{\bar z_1} h \eqno(17)
$$
This can be easily solved for $H.$ The Fourier transform of (15),
after Wick rotation to Minkowski space, with $2k_{\bar z} \rightarrow
k\cdot Q,~2k_z \rightarrow k\cdot Q'$, gives
$$
\Gamma_{\mu \nu}^{ab} = \delta^{ab}{C T^2\over 12\pi}
( 4 \pi \delta_{\mu 0} \delta_{\nu 0} - f_{\mu \nu})
\eqno(18) $$
where $f_{\mu \nu}
=\int d \Omega ~Q_\mu Q_\nu {k_0 \over k \cdot Q}.$

The three-point function involves the factor
$(\bar z_{12} \bar z_{23} \bar z_{31})^{-1} $ in addition to the
transverse $\delta$-function and color and $Q_\mu$ factors.
Using the splitting
$${1 \over \bar z_{12} \bar z_{23} \bar z_{31}} =
{ 1 \over (\bar z_{12})^2} ( {1 \over \bar z_{13}} - {1 \over
\bar z_{23} }) \eqno (19)$$
the Fourier transform can be evaluated by the same method as in
(16,17) to obtain
$$
\Gamma_{\mu \nu \lambda}^{abc} =~ f^{abc}~{iCT^2\over 12\pi}
\int d \Omega ~Q_\mu
Q_\nu Q_\lambda ~{1\over k_3 \cdot Q}
( { k_{20} \over k_2 \cdot Q} - {k_{10} \over
k_1 \cdot Q} ) \eqno(20)
$$
Expressions (18,20) agree with the diagrammatic evaluation of
hard thermal loops $^{7,11}$. We have checked the
four-point function in a similar way. For this and for the higher
point functions, a splitting formula analogous to (19) is very useful.
It is given by the Ward identity or in other words, by
the recursive buildup of the correlator of currents in the WZW model.
Combined with the Fourier transform method we have used, this gives an
efficient way of calculating the higher point functions.

The vectors $Q, Q^\prime ,$ in terms of the coordinates $u,v,$
define a two-dimensional subspace in spacetime. Our results
indicate that at high temperature, as far as the hard thermal
loops are concerned, the dynamics is essentially the C-S  dynamics
for the components $A_u, A_v.$ The final results, of course,
do not depend on the choice of this subspace since we integrate
over the orientations of $\vec Q$. The choice of this subspace
and the integration over the orientations of $\vec Q$ can be
incorporated into the action by using, instead of (5),
$$
S = \int d\Omega ~{k \over 4\pi}\int\limits_{R^5}d^5x~ {\rm Tr}
(a_\mu \partial_\nu a_\alpha +{\textstyle {2\over 3}}a_\mu a_\nu
a_\alpha)~\omega_{\beta \tau}~\epsilon^{\mu\nu\alpha \beta\tau} \eqno(21)
$$
where $\omega_{\mu\nu} = {1 \over 2} (m_\mu n_\nu -m_\nu n_\mu)$,
with $m_\mu , n_\nu$ defining a basis for vectors transverse
to the $Q-Q^\prime$-plane. This action is similar to
the K\"ahler-Chern-Simons (KCS) action considered in reference 5.
The difference is that for us $\omega$, being restricted to
directions transverse to the $Q-Q^\prime $-plane, is degenerate.
As for the KCS theory, the equations of motion tell us that the fields
do not depend on the extra fifth dimension in the action (21).

Since the C-S action is odd under parity, its presence in a QCD
calculation may be potentially worrisome. However, we do not have
any parity violation because all our results are integrated over the
orientations of $\vec Q$. Only the parity preserving contributions to
the $n$-point functions survive this integration.
Also for the non-Abelian C-S action, one has to address the issue of
quantization of the coefficient of the action.
Again, the integration over the orientations of $\vec Q$ shows that
there is no quantization. The quantization arises, in the usual
analysis, from the requirement of invariance under homotopically nontrivial
gauge transformations. In our case, there are no nontrivial gauge
transformations consistent with the angular symmetry imposed by the
integration over the orientations of $\vec Q$.
In other words, the relevant winding number corresponding to
maps of the three-sphere into the group cannot be defined in a way
that is invariant under the $\vec Q$-integration.

We conclude by rewriting the effective action in another way. Defining
$a_{\bar z}={1\over 2}Q'\cdot A,~a_z={1\over 2}Q\cdot A$, where $a_{\bar
z}$ is no longer related to $a_z$ by equation (6), we can write
$$
\Gamma= 8\pi \int{{d^3 Q}\over {(2\pi )^3}} {1\over
2Q}N(Q)~K[a_z,a_{\bar z}]
$$
$$
K[a_z, a_{\bar z}]~=~ -2C \left[ {1\over \pi}\int d^4x~{\rm Tr}(a_z a_{\bar
z})~+~
iI~+~i{\tilde I}\right]\eqno(22)
$$
Here $N(Q)$ is the Bose-Einstein distribution and
${\tilde I}$ is given by $I$ (from (12)) with the change $a_z\rightarrow
a_{\bar
z},~
z\rightarrow {\bar z}$. The density $K$ is gauge invariant; apart from
integration over the tranverse coordinates, it can be written as
${\rm Tr~log}(D_z D_{\bar z})$ where $D_z,~D_{\bar z}$ are the corresponding
two-dimensional covariant derivatives in the adjoint representation.$^{12}$
It can also be considered as the K\"ahler potential associated to the
symplectic structure for the C-S action.$^{3,4}$
Following reference 13, it may also be possible to relate this to the
forward scattering amplitude for high energy gluons on a gauge field
background.
These issues will be discussed in more detail elsewhere.
\vskip .2in
We thank A.Mueller, J.Schiff and E.Weinberg for discussions.
\vskip .3in
\noindent{\bf References}
\vskip .1in
\item
{1.} R.Jackiw and S.Templeton, {\it Phys.Rev.} {\bf D23}, 2291 (1981);
J.Schonfeld, {\it Nucl.Phys.} {\bf B185}, 157 (1981); S.Deser, R.Jackiw
and
S.Templeton, {\it Phys.Rev.Lett.} {\bf 48}, 975 (1982); {\it Ann.Phys.}
{\bf
140}, 372 (1982).
\vskip .1in
\item
{2.} See for example, S.Forte, {\it Quantum Mechanics and Field Theory
with Fractional Spin and Statistics}, Saclay Preprint SPhT/90-180,
December 1990 (to be published in {\it
Rev.Mod.Phys.});
R.Jackiw, {\it Topics in Planar Physics}, MIT Preprint CTP-1824 (December
1989).
\vskip .1in
\item
{3.} E.Witten, {\it Comm.Math.Phys.} {\bf 121}, 351 (1989).
\vskip .1in
\item
{4.} M.Bos and V.P.Nair, {\it Phys.Lett.} {\bf B223}, 61 (1989); {\it
Int.J.Mod.Phys.} {\bf A5}, 959 (1990); S.Elitzur, G.Moore, A.Schwimmer
and
N.Seiberg, {\it Nucl.Phys.} {\bf B326}, 108 (1989); J.M.F.Labastida and
A.V.Ramallo, {\it Phys.Lett.} {\bf B227}, 92 (1989); H.Murayama, {\it
Z.Phys.}
{\bf C48}, 79 (1990); A.P.Polychronakos, {\it Ann.Phys.} {\bf 203}, 231
(1990);
T.R.Ramadas, I.M.Singer and J.Weitsman, {\it Comm.Math.Phys.} {\bf 126},
409
(1989); G.V.Dunne, R.Jackiw and C.A.Trugenberger, {\it Ann.Phys.} {\bf
149}, 197 (1989).
\vskip .1in
\item
{5.} V.P.Nair and J.Schiff, {\it Phys.Lett.} {\bf B246}, 423 (1990);
Columbia
University Preprint CU-TP-521 (to be published in {\it Nucl.Phys.B}).
\vskip .1in
\item
{6.} R.Jackiw and E.Weinberg, {\it Phys.Rev.Lett.} {\bf 64}, 2234 (1990);
R.Jackiw, K.Lee and E.Weinberg, {\it Phys.Rev.} {\bf D42}, 3488 (1990);
J.Hong,
Y.Kim and P.Y.Pac, {\it Phys.Rev.Lett.} {\bf 64}, 2230 (1990).
\vskip .1in
\item
{7.} R.Pisarski, {\it Physica} {\bf A158}, 246 (1989); {\it
Phys.Rev.Lett.} {\bf
63}, 1129 (1989); E.Braaten and R.Pisarski, {\it Phys.Rev.} {\bf D42},
2156
(1990); {\it Nucl.Phys.} {\bf B337}, 569 (1990); {\it ibid} {\bf B339},
310
(1990); Brookhaven-Northwestern Preprint BNL 46, NUHEP-91-21 (November
1991, to be published in {\it Phys.Rev.}{\bf D}).
\vskip .1in
\item
{8.} J.C.Taylor and S.M.H.Wong, {\it Nucl.Phys.} {\bf B346}, 115 (1990);
J.Frenkel and J.C.Taylor, DAMTP Preprint, December 1991.
\vskip .1in
\item
{9.} B.M.Zupnik, {\it Phys.Lett.} {\bf B183}, 175 (1987).
\vskip .1in
\item
{10.} E.Witten, {\it Comm.Math.Phys.} {\bf 92}, 455 (1984).
\vskip .1in
\item
{11.} J.Frenkel and J.C.Taylor, {\it Nucl.Phys.} {\bf B334}, 199 (1990).
\vskip .1in
\item
{12.} A.M.Polyakov and P.B.Wiegmann, {\it Phys.Lett} {\bf 141B}, 223
(1984); D.Gonzales and A.N.Redlich, {\it Ann.Phys.} {\bf 169}, 104
(1986); D.Karabali, Q-H Park, H.J.Schnitzer and Z.Yang, {\it Phys.Lett.}
{\bf 216B}, 307 (1989); D.Karabali and H.J.Schnitzer, {\it Nucl.Phys.}
{\bf B329}, 649 (1990); K.Gawedzki and A.Kupianen, {\it Phys.Lett.}
{\bf 215B}, 119 (1988); {\it Nucl.Phys.}, {\bf B320}, 649 (1989).
\vskip .1in
\item
{13.} G.Barton, {\it Ann.Phys.} {\bf 200}, 271 (1990).
\end